\documentclass[11pt,twoside]{article}
\usepackage{./asp2014}

\aspSuppressVolSlug
\resetcounters

\begin{document}

\title{Science with an ngVLA: Debris disk structure and composition}
\author{B.C.~Matthews,$^{1,2}$ J.S.~Greaves,$^3$ G.~Kennedy$^{4,5}$, L. Matr\`a$^6$, D.~Wilner$^6$, and M.C.~Wyatt$^5$
\affil{$^1$National Research Council of Canada, Herzberg Astronomy \& Astrophysics Research, Victoria, BC, Canada; \email{brenda.matthews@nrc-cnrc.gc.ca}}
\affil{$^2$University of Victoria, Department of Physics \& Astronomy, Victoria, BC, Canada}
\affil{$^3$Cardiff University, Cardiff, Wales, UK}
\affil{$^4$University of Warwick, Warwick, UK}
\affil{$^5$University of Cambridge, Institute of Astronomy, Cambridge, UK}}
\affil{$^6$Harvard-Smithsonian Center for Astrophysics, Cambridge, MA, USA}

\paperauthor{Brenda C. Matthews}{Author1Email@email.edu}{ORCID_Or_Blank}{Author1 Institution}{Author1 Department}{City}{State/Province}{Postal Code}{Country}
\paperauthor{Jane S. Greaves}{Author2Email@email.edu}{ORCID_Or_Blank}{Author2 Institution}{Author2 Department}{City}{State/Province}{Postal Code}{Country}
\paperauthor{Grant Kennedy}{Author3Email@email.edu}{ORCID_Or_Blank}{Author3 Institution}{Author3 Department}{City}{State/Province}{Postal Code}{Country}
\paperauthor{Luca Matr\`a}{Author4Email@email.edu}{ORCID_Or_Blank}{Author4 Institution}{Author4 Department}{City}{State/Province}{Postal Code}{Country}
\paperauthor{David Wilner}{Author5Email@email.edu}{ORCID_Or_Blank}{Author5 Institution}{Author5 Department}{City}{State/Province}{Postal Code}{Country}
\paperauthor{Mark C. Wyatt}{Author6Email@email.edu}{ORCID_Or_Blank}{Author6 Institution}{Author6 Department}{City}{State/Province}{Postal Code}{Country}

\begin{abstract}
Debris disks, comprised of planetsimal belts and the dust and gas produced by their mutual collisions, are the longest-lived phase of circumstellar disks. Typically much fainter in emission than protoplanetary disks, debris disks can be found in associations in which some members still host protoplanetary disks rich in gas as well as around main sequence stars of all ages. There are even classes of debris disks around AGB stars and white dwarfs. Typically, these disks have been studied through dust emission though an increasing number of young disks are now found to host some gas which may be remnants of the protoplanetary disks or second-generation gas. The ngVLA will have a particular niche in the study of the dust population of these disks, since the size distribution of the dust can be derived from the spectral index of the spectral energy distribution from the far-infrared to the centimetre.  With the longest lever arms provided by ngVLA, the size distribution can be characterized, testing models of collisional evolution. 
The ngVLA could also be capable of measuring the quantities of HI and OH emission associated with the disks. HI and OH are by-products of the dissociation of water, which is expected to be a primary component of outgassed material from cometary collisions. A probe of the water content of extrasolar systems has important implications for the delivery of water to terrestrial, potentially Earth-like planets. 
\end{abstract}

\section{Introduction}

Debris disks are collisionally-dominated circumstellar disks comprised of planetesimal belts of solid bodies of 100s to 1000s of km in
size. These collisionally evolve to populate circumstellar dust and gas disks that may exist co-spatially with the planetesimal bodies,
but also may be subject to radiation and drag forces which act to
remove small grains and gas from the systems \citep[e.g.,][]{hughes2018,matthews2014,wyatt2008}. Detectable around stars from very young
ages (a few Myr) to evolved stars of many Gyr, debris disks represent
the longest-lived phase of circumstellar disks. Their presence is an
indication that planet formation processes in the system were successful to point of
producing the parent bodies of the observed dust belt. The timescales for
removal of the observed dust grains in all cases is greatly exceeded by the
age of the star, meaning the dust must be continuously replenished through a
collisional cascade from the largest sized objects to the dust grains.

The presence of bright debris
disk emission can serve as a signpost for planetary mass bodies and
orbital evolution of those bodies, which can stir the planetesimal
disk and lead to enhanced collisional rates, even at very late ages.
Observations of debris disks at mm and cm wavelengths have an
advantage over optical scattered light imaging or thermal imaging at infrared
wavelengths because the mm and cm observations probe the largest dust
grains that can be detected in the disk, and they are most likely to be
found near their place of creation (i.e., in the planetesimal belts),
since they are not as easily removed from the systems through radiative or
stellar wind forces as the smaller dust grains.  They can also trace, through modeling, the
underlying mass of the planetesimal belts which cannot be measured directly in extrasolar systems.

Debris disks are fundamental parts of planetary systems and many
systems show evidence of multiple planetesimal belts, akin to the Solar System's own asteroid and cometary belts.  Debris
disk material can interact with the planets in a system, creating
recognizable substructure in the disk due to resonances or secular
orbital evolution.  Thus, debris disks offer an opportunity to detect
evidence of planets of relatively low mass at relatively large orbits (for example, a Neptune mass planet at Neptune's orbit) around stars as distant as the Sco-Cen association at 125 pc 
which are not
detectable by any other planet detection technique.  Even in cases
where no significant asymmetry is observed, the inner edge of the disk
varies in width and eccentricity with the proximity and mass of the
nearest planet mass, and so high resolution imaging of disks can
constrain the presence and characteristics of a planet in such systems
\citep[e.g., $\eta$ Corvi][]{marino2018}.

\section{Goals and Scientific Importance}

\subsection{The size distribution of dust in debris disks}

One of the key components of a debris disk is the population of dust grains produced through the collisional evolution of planetesimals. The planetesimals link the debris disks to the planet formation process, but they themselves are not observable. Instead, we must glean information about the collisional evolution of the disk through measurements of the mass in dust grains produced, the distribution of the scattered light or thermal emission, and the spectral energy distribution, which reveals the temperature of the dust grains (and therefore their minimum radial separation from the star), the dominant size of dust grains present, the size distribution of those grains and, in the cases of mid-IR and other spectroscopy, the composition of the grains themselves.   The clearest window into the size distribution is the spectral slope between the far-infrared and cm, which reveals whether the dust component of the disk has been generated in a collisional cascade or a more stochastic process, such as a recent collision in the disk.  This latter explanation may be responsible for some of the observed bright debris disks around relatively old stars. 

In the mm/submm, ALMA provides a means of measuring dust populations down to relatively low masses, dependent on our understanding of the emissivity of the grains. With ALMA, we can also measure the spectral slope in the mm/submm through measurements in multiple bands.   Due to the emissivity of individual grains, the mm data do not sample cm-scale grain sizes. Whether the spectral slope is constant or turns over in the cm also has implications for the collisional evolution of the disk. Measurement of the size distribution toward a significant number of disks (especially at young ages) will provide important insight into their (early) evolution. To sample the youngest disks, we need sensitivity to their emission in nearby star forming regions, such as Sco-Cen ($d = 125$ pc).

It is not the case that all observed disks present a classical Dohnanyi (1969) profile for their size distribution, which is a power law of the form $dN/da \propto a^{-q}$ where $q=3.5$.  There is in fact a range of measured values for $q$ between 3 and 4 through numerical simulations \citep[e.g.,][]{pansari2005,panschlichting2012,gaspar2012}. \cite{draine2006} showed that this value can be derived directly from the long wavelength slope of SEDs, so coverage throughout the range of the far-IR and cm improves our measure of $q$.  Coverage into the cm regime could reveal whether there are any kinks in this distribution for any size scale to cm size grains. 

Resolved imaging at multiple wavelengths would reveal a radially resolved spectral index in the mm/cm regime. These spectral indices can be different if the debris is fed by different types of planetesimals or different collisional processes, such as a steady state collisional cascade versus a giant planet impact in the disk (e.g. HD 111520, Draper et al., 2018, in prep). The spectral index is different for fluffy versus icy particles. 

When sources are well resolved from the star, it is also possible to isolate the stellar emission from the disk, eliminating a potentially complicating factor at these wavelengths, since the stellar photosphere is no longer the dominant source of stellar emission \citep[e.g.,][]{cranmer2013}.  The chromosphere of the star produces a rise in emission in the mm and cm that can mimic the effects of a debris disk (i.e., an excess above the level of the photospheric emission extrapolated from the far-IR). Therefore, to interpret the debris disk emission properly, the resolution should be sufficient to separate the stellar emission spatially from the disk emission, which is typically ~ 40 AU - 100 AU, driving us to resolutions of at least 1\arcsec, but with good sensitivity to scales out to 100s of AU.

When azimuthal structure is present in a disk, the collision rate changes as a function of both radius and azimuth in the disk. Two scenarios for producing this type of structure are i) long-term perturbations from an eccentric planet, which offsets the center of the disk from the star, and ii) capture of planetesimals into resonances during outward planet migration, which creates overdensities (clumps) of material that rotate with the planet. The disk structure resulting from these scenarios is strongly wavelength dependent \citep{wyatt2006} for two reasons: the first is the aforementioned differences in radiation forces as a function of grain size, and the second is that the collision rate is now a function of location in the disk. Observations at a wide range of wavelengths, especially from the mm-cm, is diagnostic of the size distribution, where collisions are prevalent, and the likely nature of the (probably unseen) perturber.  

While there are only a few detections of debris disks in continuum emission at cm wavelengths to date \citep[e.g.,][]{greaves2012,ricci2012,ricci2015,macgregor2016_tauceti,marshall2017}, the ngVLA would significantly expand the potential for this science in the centimeter regime. The real challenge is sensitivity, and with enhanced sensitivity, nearby disks could be imaged in a range of wavelengths. In concert with ALMA, which has now produced well resolved images of about two dozen debris disks (and counting!), these data could probe the size distribution of the solids within the disks. The effect of differing dust temperature is mitigated at longer wavelengths.  The spatial distribution of grains of different sizes will be measurable by comparing resolved imaging from the ngVLA and ALMA. 

\begin{figure}
\begin{center}
\includegraphics[scale=0.25]{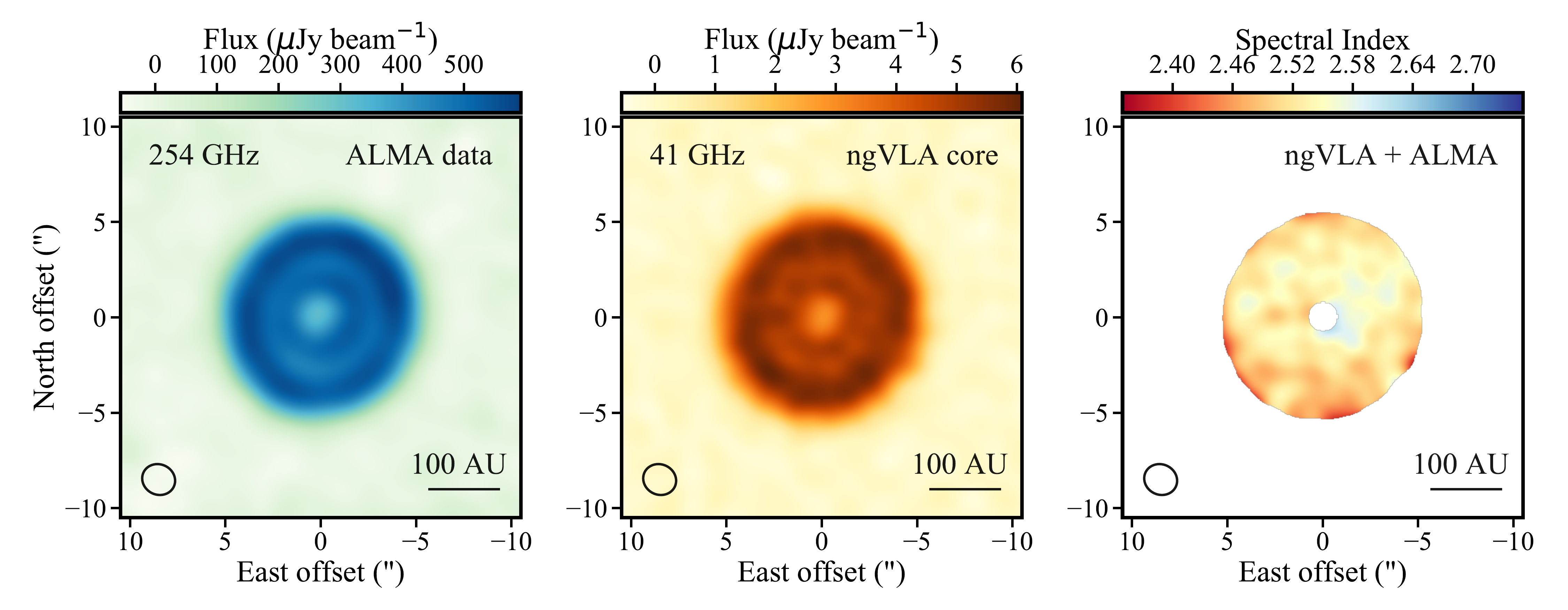}
\includegraphics[scale=0.25]{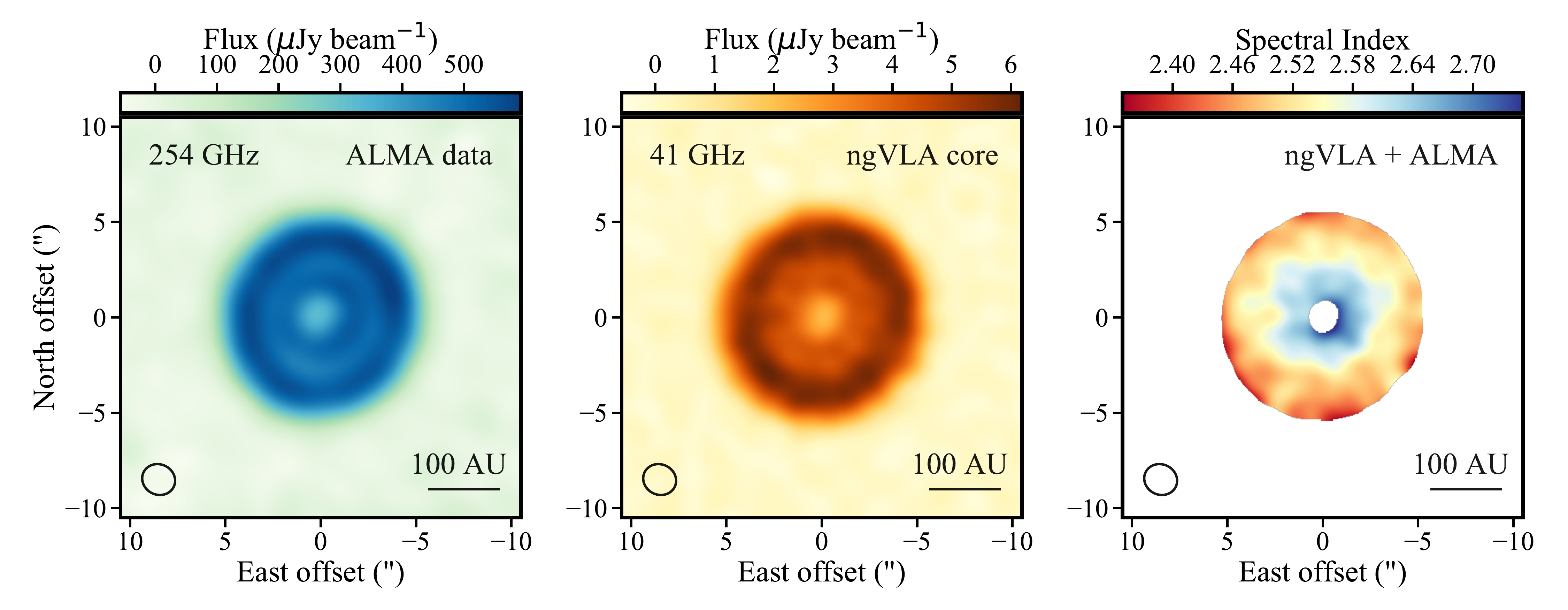}
\caption{
ALMA 1.2 mm observations \citep{marino2018} and ngVLA 7 mm simulations of the
debris disk around HD 107146, a 100 Myr-old Solar analog (spectral type G2V) at d = 27.5 pc.
The gap in the disk at a radius of 80 AU may be caused by the dynamical influence of an unseen planet.
The two rows illustrate two scenarios, one where the spectral index is invariant with position (upper
row), and one where the spectral index changes with radius between 20 to 170 AU by an amount that
corresponds to a decrease of 0.15 in $q$, the power-law index of the grain size distribution (lower row).
For each scenario we show: {\it (Left)} ALMA band 6 image smoothed to match the 1.67 arcsec (46 AU) resolution of
the ngVLA imaging simulation; {\it (Center)} Simulated ngVLA band 5 image using the 114 core antennas, at
41 GHz, with 10 h integration time, where the disk total flux density was extrapolated from
mm/submm data using the observed spectral index of 2.6); {\it (Right)} Resolved spectral index map of the
emission obtained using ALMA and the ngVLA (masked where signal-to-noise ratio falls below 20).
The simulated ngVLA observations clearly recover the radial dependence in the spectral index the
lower row, easily distinguishable from the simulation with no radial dependence in the upper row.
}
\label{debrisdisksrock}
\end{center}
\end{figure}

There are at least 50 such systems in a feasible brightness range within 100pc, though not all of these will show detailed structure that requires the highest spatial resolution. A typical observing strategy would be to initially image tens of systems at relatively low resolution to identify the subset that exhibit asymmetries, and then dedicate significant effort to characterizing that subset. Simulations (see Figure \ref{debrisdisksrock}) show that we will be able to discriminate relatively modest changes in spectral slope with a combination of ALMA and ngVLA data. The lower resolution imaging will yield broad-brush measures of disk sizes as a function of wavelength, and disk-integrated mm spectral slopes to study the dust size distribution.

To address this science, we would require the measurement of flux densities over a range in wavelengths (coordinated with the mm/submm) toward a significant numbers of targets. The goal would not be to resolve each of the disks, but get a bulk measure of the flux density at each wavelength. In nearby star-forming regions, the numbers of detected debris disks number in the 10s, with 100s of potential targets in total, depending on the sensitivity limits.  In the mm, debris disks around nearby stars, such as those in the 50 pc TW Hya association have flux densities in the ~ 1 mJy range at ALMA Band 6. There is variation, but this is a reasonable flux density expected in ALMA Band 6. Typically, we observe the flux density falling off with a $\beta$ of 1, $\alpha$ = 3 in the R-J limit, where $S_\nu \propto \nu^{-\alpha}$ and $\alpha = 2 + \beta$. This implies expected flux densities in the 3mm band (100 GHz) to be 10s of uJy, on the edge of observable with ALMA (2 uJy/beam rms in Band 3 requires > 1 day integration time with ALMA, but reachable in 1 hour of integration time with the ngVLA.  At longer wavelengths, such as 7 mm, the flux densities will be even lower, on the order of ~1 uJy/beam.  This would require rms levels on the order of a fraction of a uJy/beam (0.2 uJy for a 5 sigma detection).  Not all disks would necessarily be this faint, but reaching this level of rms would increase the number of viable disk candidates.  Rms values as low as 0.2 uJy would be required, which are achievable in ~5 hours with the ngVLA, but not with Band 1 with ALMA. For this science, we would not require the highest resolutions achievable with ngVLA; in fact, we will not be concerned with resolution in this project at all. Most debris disks are no larger than ~40-100 AU in radius, and at 50 pc, this implies a 2-4" beam will suffice for this science without resolving the disk emission. Baselines of 0.2 km will be more than sufficient to probe these objects (3" at 100 GHz, 1.2" at 40 GHz), resulting in good beam matching to the targets at 50 pc. For closer targets, a larger beam may be needed to prevent surface brightness considerations from being taken into account.

\subsection{Presence and nature of gas}

Debris disks were once thought to be essentially devoid of gas, but an increasing number of young disks around A and F stars have detected gas, typically CII or CO \citep[for a detailed review, see][]{hughes2018}.  Around young stars particularly, the nature of the detected gas is ambiguous: is it collisionally produced like the dust, or is the gas a remnant of the putative protoplanetary disk?  With the ngVLA, we can detect atomic hydrogen (HI) through its 21 cm (1.42 GHz) line in nearby debris disks.  These observations will establish whether HI is created from photodestruction of water released from colliding exocomets, or a whether it is remnant of the protoplanetary phase of planet formation. In the exocometary scenario, the HI measurement will be compared to ongoing measurements of carbon-bearing species (CI, CII, CO with Herschel/ALMA) to estimate the C/H ratio and water content of exocomets at the epoch where volatile delivery events are most likely to take place.

Establishing the water content of planetesimals has far reaching implications for the understanding of how and how much water is delivered to terrestrial planets.  Detection of water from comet outgassing in the solar system is typically done via measurements of OH, a dissociative product of water under solar radiation. Debris disks are created by the collisional destruction of planetesimal bodies; as such, they can provide a window into the water content of planetesimals around other stars. Unlike the solar system, there is no way to observe individual planetesimals within debris disks.  The importance of water for the evolution of life as we know it is well known; determining whether the fraction of water in extrasolar comets is comparable to those in the solar system would provide evidence of how much variation there is in the water delivery systems in other solar systems. 

Direct detection of $H_2O$ is prohibitively difficult due to its short photodissociation timescale, but its products OH and HI provide a means of determining how much water could be present in planetesimals. OH provides the most direct probe, since one OH molecule is produced for each $H_2O$ molecule dissociated, but similarly to water should be readily photo-destroyed in typical debris disk environments. Therefore, neutral hydrogen (HI) produced via the first dissociation of water and the subsequent dissociation of OH is the most promising probe of the H2O content of extrasolar planetesimals.  In addition, there remains a debate on whether gas-bearing debris disks around very young systems may instead be tenuous remnants of the protoplanetary disks, which would have significant implications on the lifetime of protoplanetary disks and the formation of gas giant planets. In this scenario, HI is readily produced by photodestruction of $H_2$ gas due to UV and/or X-ray radiation easily penetrating through an increasingly low mass, and hence increasingly optically thin gas disk.  Detection of low levels HI - as would be enabled by the ngVLA - is therefore of paramount importance to either estimate the water content of exocomets or to assess the behavior of the bulk of the gas mass in a dispersing protoplanetary disk, as well as to distinguish between these two scenarios.

Deep, high-sensitivity observations of HI at 1.42 GHz are required for detection. To optimize the latter, we require a spectral resolution equal to the expected line width (few-few tens km/s depending on system viewing geometry), and a spatial resolution of order the maximum expected disk diameter (1"-10" for typical systems between 10 and 100 pc). We do not require imaging capability; resolution of the disks would require greatly enhanced sensitivity due to the low surface brightness of the emission. For the unresolved case, the sensitivity required is ~50 uJy/beam per 2 km/s channel for a 2" beam to be able to achieve unresolved detection of 0.08 $M_{Earth}$ of HI produced from exocometary water in a planetary system at 50 pc, assuming a typical belt location at ~50 AU distance from the star viewed face-on. Models of exocometary gas release \citep{matra2015,kral2017_gas} predict $\sim 4$ systems to be detectable at a significance $> 3\sigma$. If the gas (already detected in CO for 11 disks) is a tenuous remnant of the protoplanetary disk, the generally higher HI masses predicted \citep[in the range ~1-10 $M_{Earth}$][]{kamp2007} would be detectable at high significance in systems as far as $\sim 100$ pc.

\section{Limitations of current instrumentation}

This science case requires resolved imaging of the dust emission in the debris disk with sufficient sensitivity to measure of the spatial distribution of dust emission. To establish the dust size distribution, the disks must be resolved at multiple wavelengths. Where possible and necessary, the disks should be resolved over tens of beams to quantify the azimuthal structure at multiple wavelengths, to measure the size distribution, identify where dust is being produced in collisions, and map out the parent planetesimal locations. Frequencies from 30-80 GHz provide good sensitivity to thermal dust emission from large grains, the sensitivity at 30 GHz will the limiting factor in what can be imaged. An extension to 100 GHz would be useful to better define the spectral shape. 

As noted above the super-sample of probable targets contains about 50 debris disk systems within 100 pc. Typical detections at 1mm are about 1-10mJy, corresponding to 10-100 uJy at 4 mm and 1-10 uJy at 1 cm. A subset of the nearest systems may not be amenable for ngVLA because their spatial extent is too large, but most have sizes from 1-10". Typically disks have a relatively narrow ring structure, so the disk surface brightness scales with radius (not r squared), and therefore the desired resolution is roughly 0.01-1\arcsec, depending on the specific brightness, structure, and geometry of the system in question (which may be estimated from observations at other wavelengths or from lower resolution reconnaissance). To study the spatial structure requires several tens of beams around the disk; assuming 30 beams at a S/N of 3 therefore dictates a sensitivity of 0.01-0.1 uJy/beam at 1 cm (this is the most stringent requirement in terms of sensitivity, imaging at shorter wavelengths to the same sensitivity would require less time).

\section{Uniqueness of ngVLA}

Imaging debris disks requires very high sensitivities as the disks have fractional luminosities significantly below those of proto-planetary disks (i.e., $< 10^{-2}$, down to Solar levels of $10^{-7}$, which are not yet detectable; ratios of $10^{-5}$ are typical of detected disks to date). The masses of debris disks in dust grains are also low (several to a fraction of a lunar mass), and dust opacities drop steeply with wavelength. For this reason, very few debris disks have been detected longward of 3 mm \citep[e.g.,][]{ricci2012,marshall2017}. However, the increased sensitivity of the ngVLA will make it competitive for the detection of disks detected by ALMA at Band 6 and 7 at wavelengths of 3 mm, which ALMA cannot achieve in Band 3. This will permit a probe of the cm-scale solids in the disk, which are expected to hold significantly more mass than the mm grains. Measuring the size distribution from the spectral slope in the mm/cm regime will reveal how the size distribution varies in the disk, which will permit testing of collisional models.

No other facility will be able to address this science. The Jansky VLA is not sensitive enough to probe debris disks effectively excepting the brightest nearby ones. The sensitivities are such that continuum emission from debris disks, even at $\sim 50$ pc, could be detectable to the notional model of the ngVLA in timescales of hours at 42 and100 GHz, which is unachievable to ALMA at these wavelengths.  The SKA-1 doesn't approach these observing frequencies. The very nearby debris disks, while brighter, will have different requirements, since they are more likely to be resolved (see separate science case).

The ngVLA is the only observatory that can achieve the required high sensitivity to detect low HI masses in emerging planetary systems at 1.42 GHz, and this is the only way to measure the total hydrogen content in gas-bearing debris disks. In the exocometary release scenario, the ngVLA would be the only facility allowing us for the first time to probe the water content of exocomets, allowing comparison with the current-day Solar System comets and how they used to be in the epoch of water delivery to the Earth.

\section{Complementarity}

The measurements of the size distribution in debris disks cannot be done solely with the ngVLA, but nor can it be done by any other individual instrument. To map out the disk structure as a function of grain size requires disk observations over a wide range of wavelengths, from the optical to cm, all at a sufficient resolution to detect the disk over tens of beams. Thus, instruments that can image disks, such as ALMA, JSWT, WFIRST, OST, and LUVOIR all have a role to play. ALMA of course will remain a highly effective instrument for the study of these disks in Bands 6 and 7, but most are be too faint to obtain effective observations out to 50 pc at 3mm and 7mm (Bands 3 and 1).  The unique contribution of the ngVLA is that it is best suited to detecting the particles least affected by radiation forces at 3, 7 and 10 mm, so will yield the spatial distribution of the parent planetesimals, from which the dust detected at all other wavelengths ultimately originates.

Observations of HI with the ngVLA will be complemented by the SKA for targets in the Southern sky, completing the all-sky sample. As well, HI observations will be combined with CII+OI observations available from Herschel and planned with the increased sensitivity of the Origins Space Telescope, as well as ALMA CO+CI observations to reach a complete view of the volatile composition of extrasolar planetesimals (in the exocometary scenario) or to explore how the chemical composition of a protoplanetary disk evolves as it disperses (in the primordial scenario). Finally, JWST will also provide valuable information on the disks' molecular content through detection of ro-vibrational transitions.

\acknowledgements ...  

\bibliography{ngVLA_debrisdisks}  

\begin{thebibliography}{}
\expandafter\ifx\csname natexlab\endcsname\relax\def\natexlab#1{#1}\fi
\expandafter\ifx\csname url\endcsname\relax
  \def\url#1{\texttt{#1}}\fi
\expandafter\ifx\csname urlprefix\endcsname\relax\def\urlprefix{URL }\fi
\providecommand{\eprint}[2][]{\url{#2}}

\bibitem[{{Cranmer} et~al.(2013){Cranmer}, {Wilner}, \&
  {MacGregor}}]{cranmer2013}
{Cranmer}, S.~R., {Wilner}, D.~J., \& {MacGregor}, M.~A. 2013, \apj, 772, 149.
  \eprint{1306.4567}

\bibitem[{{Draine}(2006)}]{draine2006}
{Draine}, B.~T. 2006, in Astrophysics in the Far Ultraviolet: Five Years of
  Discovery with FUSE, edited by G.~{Sonneborn}, H.~W. {Moos}, \& B.-G.
  {Andersson}, vol. 348 of Astronomical Society of the Pacific Conference
  Series, 58

\bibitem[{{G{\'a}sp{\'a}r} et~al.(2012){G{\'a}sp{\'a}r}, {Psaltis}, {Rieke}, \&
  {{\"O}zel}}]{gaspar2012}
{G{\'a}sp{\'a}r}, A., {Psaltis}, D., {Rieke}, G.~H., \& {{\"O}zel}, F. 2012,
  \apj, 754, 74. \eprint{1111.0296}

\bibitem[{{Greaves} et~al.(2012){Greaves}, {Hales}, {Mason}, \&
  {Matthews}}]{greaves2012}
{Greaves}, J.~S., {Hales}, A.~S., {Mason}, B.~S., \& {Matthews}, B.~C. 2012,
  \mnras, 423, L70

\bibitem[{{Hughes} et~al.(2018){Hughes}, {Duchene}, \& {Matthews}}]{hughes2018}
{Hughes}, A.~M., {Duchene}, G., \& {Matthews}, B. 2018, ArXiv e-prints.
  \eprint{1802.04313}

\bibitem[{{Kamp} et~al.(2007){Kamp}, {Freudling}, \& {Chengalur}}]{kamp2007}
{Kamp}, I., {Freudling}, W., \& {Chengalur}, J.~N. 2007, \apj, 660, 469.
  \eprint{astro-ph/0701248}

\bibitem[{{Kral} et~al.(2017){Kral}, {Matr{\`a}}, {Wyatt}, \&
  {Kennedy}}]{kral2017_gas}
{Kral}, Q., {Matr{\`a}}, L., {Wyatt}, M.~C., \& {Kennedy}, G.~M. 2017, \mnras,
  469, 521. \eprint{1703.10693}

\bibitem[{{MacGregor} et~al.(2016){MacGregor}, {Lawler}, {Wilner}, {Matthews},
  {Kennedy}, {Booth}, \& {Di Francesco}}]{macgregor2016_tauceti}
{MacGregor}, M.~A., {Lawler}, S.~M., {Wilner}, D.~J., {Matthews}, B.~C.,
  {Kennedy}, G.~M., {Booth}, M., \& {Di Francesco}, J. 2016, \apj, 828, 113.
  \eprint{1607.02513}

\bibitem[{{Marino} et~al.(2018){Marino}, {Bonsor}, {Wyatt}, \&
  {Kral}}]{marino2018}
{Marino}, S., {Bonsor}, A., {Wyatt}, M.~C., \& {Kral}, Q. 2018, \mnras.
  \eprint{1806.01289}

\bibitem[{{Marshall} et~al.(2017){Marshall}, {Maddison}, {Thilliez},
  {Matthews}, {Wilner}, {Greaves}, \& {Holland}}]{marshall2017}
{Marshall}, J.~P., {Maddison}, S.~T., {Thilliez}, E., {Matthews}, B.~C.,
  {Wilner}, D.~J., {Greaves}, J.~S., \& {Holland}, W.~S. 2017, \mnras, 468,
  2719

\bibitem[{{Matr{\`a}} et~al.(2015){Matr{\`a}}, {Pani{\'c}}, {Wyatt}, \&
  {Dent}}]{matra2015}
{Matr{\`a}}, L., {Pani{\'c}}, O., {Wyatt}, M.~C., \& {Dent}, W.~R.~F. 2015,
  \mnras, 447, 3936. \eprint{1412.2757}

\bibitem[{{Matthews} et~al.(2014){Matthews}, {Krivov}, {Wyatt}, {Bryden}, \&
  {Eiroa}}]{matthews2014}
{Matthews}, B.~C., {Krivov}, A.~V., {Wyatt}, M.~C., {Bryden}, G., \& {Eiroa},
  C. 2014, Protostars and Planets VI, 521. \eprint{1401.0743}

\bibitem[{{Pan} \& {Sari}(2005)}]{pansari2005}
{Pan}, M., \& {Sari}, R. 2005, Icarus, 173, 342. \eprint{astro-ph/0402138}

\bibitem[{{Pan} \& {Schlichting}(2012)}]{panschlichting2012}
{Pan}, M., \& {Schlichting}, H.~E. 2012, \apj, 747, 113. \eprint{1111.0667}

\bibitem[{{Ricci} et~al.(2015){Ricci}, {Maddison}, {Wilner}, {MacGregor},
  {Ubach}, {Carpenter}, \& {Testi}}]{ricci2015}
{Ricci}, L., {Maddison}, S.~T., {Wilner}, D., {MacGregor}, M.~A., {Ubach}, C.,
  {Carpenter}, J.~M., \& {Testi}, L. 2015, \apj, 813, 138. \eprint{1510.03513}

\bibitem[{{Ricci} et~al.(2012){Ricci}, {Testi}, {Maddison}, \&
  {Wilner}}]{ricci2012}
{Ricci}, L., {Testi}, L., {Maddison}, S.~T., \& {Wilner}, D.~J. 2012, \aap,
  539, L6. \eprint{1201.3383}

\bibitem[{{Wyatt}(2006)}]{wyatt2006}
{Wyatt}, M.~C. 2006, \apj, 639, 1153. \eprint{astro-ph/0511219}

\bibitem[{{Wyatt}(2008)}]{wyatt2008}
--- 2008, \araa, 46, 339

\end{thebibliography}



\end{document}